\documentclass[a4paper]{article}
\makeatletter
\def\@seccntformat#1{\@ifundefined{#1@cntformat}%
   {\csname the#1\endcsname\quad}  
   {\csname #1@cntformat\endcsname}
}
\let\oldappendix\appendix 
\renewcommand\appendix{%
    \oldappendix
    \newcommand{\section@cntformat}{\appendixname~\thesection\quad}
}
\makeatother

\usepackage[dvipdfmx]{graphicx}

\usepackage{cite}
\usepackage[normalem]{ulem}

\usepackage[usenames,dvipsnames]{xcolor} 

\usepackage{amssymb,amsfonts,amsmath}
\usepackage[top=25truemm,bottom=25truemm,left=20truemm,right=20truemm]{geometry}
\usepackage{bm}
\usepackage[mathlines]{lineno}

\usepackage{setspace}

\title{Cellular automaton model with turning behavior in crowd evacuation}
\bigskip
\author{Daiki Miyagawa${}^{1}$, Genki Ichinose${}^{1*}$
\ \\
\ \\
${}^{1}$
Department of Mathematical and Systems Engineering, Shizuoka University, \\3-5- 1 Johoku, Naka-ku, Hamamatsu, 432-8561, Japan\\
$^*$ Corresponding author (ichinose.genki@shizuoka.ac.jp)}

\begin{document}

\maketitle

\section*{Abstract}
Effective evacuation policies in emergency situations are important to save lives.
To develop such policies, simulation models based on cellular automata have been used for crowd evacuation dynamics.
In most previous studies of crowd evacuations, an evacuee is represented by a $1 \times 1$ square.
However, a rectangle ($1 \times 2$) representation is more suitable for such models than the square representation because of evacuees' shoulder width.
The rectangle representation gives two new features to evacuees' behaviors: moving sideways and turning.
We study the effects of these behaviors on crowd evacuation dynamics.
Hence, we constructed a cellular automaton model where evacuees whose shoulder widths are $1 \times 2$ try to escape from a room in an emergency situation.
The simulation results showed that turning behavior can make the evacuation time shorter and there is an optimal turning rate for the crowd evacuation.
Our findings contribute to the effective control of evacuees in emergency situations.

\section*{Keywords}
Crowd evacuation; Cellular automaton; Simulation; Turning behavior; Multi-grid model

\section{Introduction\label{sec:introduction}}
People gather in a room when working at an office, watching movies, or taking classes at school.
People gather in a larger space at mass events such as the Olympic games, fireworks, concerts, and conferences.
In those situations, when a sudden danger such as fire, an earthquake, or a terrorist attack happens, many lives may be lost if there are no effective evacuation strategies.
To develop those strategies, various crowd evacuation scenarios are needed.
These interests have recently become a popular research field \cite{Helbing2013Encyclopedia} and are important in the design of large spaces that can contain many people \cite{Helbing2005TransportSci}.

The crowd evacuation dynamics of people, namely, evacuees have been extensively investigated by theoretical simulation models \cite{Helbing2001RevModernPhys, Helbing1995PhysRevE, Tanimoto2010PhysicaA, Yuan2007PhysicaA, Helbing2000Nature, Zanlungo2011EPL, Varas2007PhysicaA, Alizadeh2011SafSci, Yang2002ChinSciBull, Zheng2011PhysicaA, Huang2008PhysRevE, Huang2015ApplMathComput, Huang2017ApplMathComput, Fu2015PhyscaA, Li2015PhysicaA, Guo2013PhysicaA, Wei2012ProcediaEng, Li2008PhysicaA, Kirchner2002PhysicaA, Henderson1971Nature, Helbing1998ComplexSyst, Chen2018SimulModelPractTheory, Ibrahim2018PhysicaA, Cheng2018PhysicaA, Bouzat2014PhysRevE, Tajima2001PhysicaA, Burstedde2001PhysicaA, Bellomo2011SIAMRev, Takimoto2003PhysicaA, Song2016PhysicaA, Tian2018PhysicaA, Cheng2018ApplMathComput, Zheng2011PhyscaA, Ma2016PhysicaA, Guo2008JPhysA}, experiments (or real data) \cite{Seyfried2009TransportSci, Georgoudas2011SystJ, Tsiftsis2016SystJ}, and models combined with experiments \cite{Liu2018SimulModelPractTheory, Ronchi2014FireSafJ, Haghani2017PhysicaA, Helbing2005TransportSci, Helbing2003PhysRevE, Guo2012TranspResB, Isobe2004PhysRevE}.
Among them, theoretical simulation models based on cellular automaton on which we focus in this paper play an important role in understanding the nature of evacuation dynamics \cite{Li2019PhysicaA}.
In recent models, researchers have included some psychological aspects during evacuation.
For example, some people tend to yield to others to exit while some other people tend to elbow people aside and may go to the exit quickly.
These behaviors are implemented as cooperation and defection strategies, respectively.
The idea comes from game theory and the effect of such strategies on evacuation dynamics has been explored \cite{Song2016PhysicaA, Bouzat2014PhysRevE, Cheng2018PhysicaA, Tian2018PhysicaA, Cheng2018ApplMathComput, Zheng2011PhyscaA}.
This requires the cognitive abilities of human beings to detect others' strategies.

Not only such high-level abilities of human beings but also the physical features of people affect crowd evacuation dynamics.
The physical features are basic rather than cognitive abilities but have not been fully investigated in the evacuation dynamics.
In the cellular automaton and lattice-gas models for crowd evacuation \cite{Kirchner2002PhysicaA, Varas2007PhysicaA, Zheng2011PhyscaA, Burstedde2001PhysicaA, Guo2008JPhysA, Guo2012PhysicaA, Song2016PhysicaA, Cheng2018PhysicaA, Fu2015PhyscaA, Cheng2018ApplMathComput, Chen2018SimulModelPractTheory, Tian2018PhysicaA, Li2008PhysicaA, Wei2012ProcediaEng, Helbing2003PhysRevE}, an evacuee is represented by a $1 \times 1$ square (or circle).
However, because of shoulder width for evacuees, the size of evacuees should be represented by a $1\times 2$ rectangle rather than the square.
In the evacuee counter flow models \cite{Muramatsu1999PhysicaA, Tajima2002PhysicaA, Weng2006PhysRevE, Takimoto2002PhysicaA, Fukamachi2007PhysicaA}, such multi-grid (or ellipse) representations ($1 \times 2$ or $m \times n$) have already been studied \cite{Fukamachi2007PhysicaA, Nagai2006PhysicaA, Jin2017PhysicaA, XuSong2009BuildEnv, Yamamoto2019TranspResPtB-Methodol} but not studied in a crowd evacuation.
Thus, we consider such effects on crowd evacuation by cellular automaton simulation models.
The rectangle type representation of an evacuee gives two new features to the dynamics:
1) The difference between moving forward and sideways and 2) turning behavior.
These features would greatly affect evacuation dynamics.
In this paper, we constructed a cellular automaton model where evacuees whose shoulder widths are $1 \times 2$ try to escape from a room in an emergency situation.
The simulation results showed that turning behavior can make the evacuation time shorter and there is an optimal turning rate for the crowd evacuation.

\section{Model\label{sec:model}}
\begin{figure}[tb]
	\centering
	\includegraphics[width=141mm]{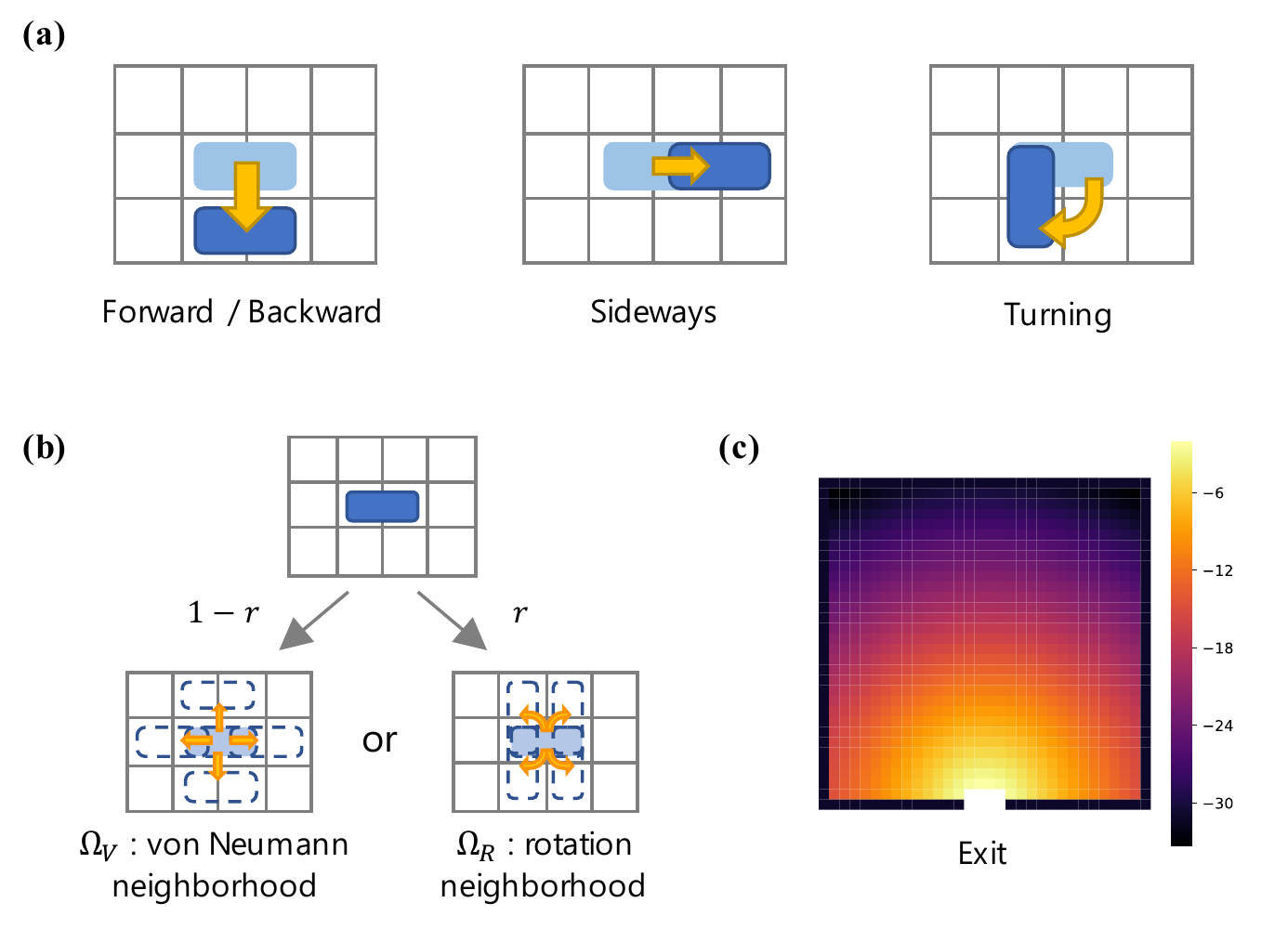}
	\caption{(a) Three types of movements. Forward/Backward movements (left): Evacuees face the exit and move toward or away. Sideways (center): Evacuees face sideways to the exit and move right or left. The speed of sideways movement is basically slow which is represented by $v$. Turning (right): Evacuees turn clockwise or counter clockwise.
	(b) Selection of the next site by turning or not. With probability $r$, evacuees turn and select one of five sites (current, back and clockwise, forward and clockwise, back and counterclockwise, and forward and counterclockwise).
	With probability $1-r$, evacuees select one of five sites (current, back, forward, right, and left).
	(c) The Static Floor Field. Evacuees roughly know the location of the exit. Thus, evacuees tend to move to brighter sites because those sites are located near the exit.}
	\label{fig_diversity}
\end{figure}

Crowd evacuation dynamics can be modeled by cellular automata.
We consider a room with a size of $L \times L$ and the room has one narrow exit with a size of $D_w = 1 \times 4$ in the bottom as seen in Fig.~\ref{fig_diversity}(c).
Basically, evacuees try to escape from the room by moving toward the exit. The color in Fig.~\ref{fig_diversity}(c) represents the intensity of moving to be selected by evacuees.
As the color is brighter, the probability to be selected becomes larger.
Each cell can be empty or contain half of an evacuee.
An evacuee occupies two adjacent cells as shown in Fig.~\ref{fig_diversity}(a).
If the form of an evacuee is represented by a $1 \times 1$ square, there is only one type of movement, which is \textit{forward} or \textit{backward}.
However, if we consider a $1 \times 2$ rectangle for an evacuee, the diversity of the movement types increase to three; \textit{forward/backward}, \textit{sideways}, and \textit{turning} (Fig.~\ref{fig_diversity}(a)).
Forward and backward movements imply that an evacuee moves toward the long side of a rectangle.
Forward movement faces the exit and backward faces the opposite.
Sideways is the movement toward the short side (right or left) of a rectangle.
Turning changes the body orientation of an evacuee.
We assume the size of an evacuee as 0.20 m $\times$ 0.40 m \cite{Cheng2018ApplMathComput}.
The initial density of evacuees is given by $\rho$.
Once a simulation starts, evacuees basically move to the exit by using the three types of movements. 
Note that we prohibit the overlapping of evacuees during the simulation. 
When all evacuees finish their evacuation, the simulation ends.
We use $L=50$ and $\rho=0.5$ unless otherwise noted.
In the next subsection, the simulation flow is described in detail. 

\subsection{Simulation flow\label{sec:model_rule}}
We explain how evacuees try to escape from the dangerous room.
Evacuees follow the steps below in order.
First, each evacuee makes a decision whether to turn or not.
This process is stochastically determined.
With probability $r$, each evacuee turns (Fig.~\ref{fig_diversity}(b)).
For turning, there are five ($2 \times 2+1$) cases in total: Clockwise or counterclockwise, front or back, and staying at the same site.
The five cases are represented by $\Omega_R$ (rotation neighborhood).
Note that even the turning is selected, evacuees still move forward or backward.

With probability $1-r$, each evacuee simply moves (Fig.~\ref{fig_diversity}(b)).
If this movement is selected, each evacuee conducts forward, backward, or sideways movement.
The five locations (current, top, bottom, right, and left) are represented by $\Omega_V$ (von Neumann neighborhood).

After the decision whether to turn or not, evacuees make their decisions for the next site from their accessible neighborhood ($\Omega_R$ or $\Omega_V$). 
Basically, we assume that evacuees know the location of the exit. Thus, evacuees tend to move toward a site at which the direct distance to the exit is short.
To realize this situation, we use the Static Floor Field (SFF) \cite{Burstedde2001PhysicaA}.
The SFF is used in other cell automaton models \cite{Kirchner2002PhysicaA, Cheng2018ApplMathComput}.
The SFF is the scalar field representing how close sites are to the exit as shown in Fig.~\ref{fig_diversity}(c); for example, the value of the SFF at the site $(x,y)$ is calculated as $-\sqrt{x^2+y^2}$ from the exit as the origin.
In our model, because an evacuee is represented by $1 \times 2$, we use the average value of the SFF in the two sites when the SFF is calculated.
The transition probability toward the site $(x,y)$ is given by the SFF as follows:
\begin{eqnarray}
	\label{eq-transition}
	p_{xy} &=& \frac{v \exp{(S_{xy}/k)}}{\displaystyle{\sum_{(i,j)\in\Omega_X}} \exp{(S_{ij}/k)}}, 
	\label{eq-v}
\end{eqnarray}
where $S_{xy}$ is the SFF value at the position $(x,y)$. $\Omega_X, X \in \{R, V\}$ denotes the set of the accessible positions including the current position except wall sites. 
The parameter $k$ is the recognition noise of evacuees of the distance to the exit.
We set $k=0.1$ which implies that the noise effect is weak and evacuees almost always move toward the site that the SFF is high.
The parameter $v$ represents the relative velocity of the evacuees movement. 
Basically, because the speed of sideways movement is slow \cite{Jin2017PhysicaA}, we set $0 < v \le 1$ when evacuees move sideways while $v=1$ when evacuees perform forward, backward, or turning.
In the simulations, $v$ is used as a probability.
That is, with probability $v$, evacuees move sideways. With probability $1-v$, evacuees do not move.

After an evacuee chooses the next site for the next time step, they check whether the site is already occupied by another evacuee.
If that site is occupied, the evacuee refrains from going there.
However, the evacuee can search for the site that they can move to again.
If this second challenge fails, the evacuee stays at the same site.
In short, evacuees have a chance to search for the possible site to move to twice \cite{Bouzat2014PhysRevE}. 

Next, all evacuees try to move to the site that they chose.
In this case, conflicts among evacuees may happen, which means that two or more evacuees want to move to the same site.
If this happens, one randomly selected evacuee can move to the site.

Then, all evacuees move to their selected sites or stay at their current site simultaneously, which means that the movement of evacuees is synchronous.
Evacuees who reach the exit (the region of $D_w$) are removed from the simulation.
The rest of the evacuees repeat the above steps until they reach the exit.
When all evacuees reach the exit, the simulation ends.

\section{Results\label{sec:result}}
We focus on how the increase of movements caused by shoulder width affect the escape time of evacuees for a crowd evacuation.
First, we reveal the effect of turning on the evacuation dynamics compared to the two extreme control settings (forward and sideways).
Then, we find the optimal value of the turning probability $r$ depending on the relative speed of sideways movement $v$.

\subsection{Comparison of escape time with and without turning\label{sec:result_comp}} 
\begin{figure}[bt]
	\centering
	\includegraphics[width=142mm]{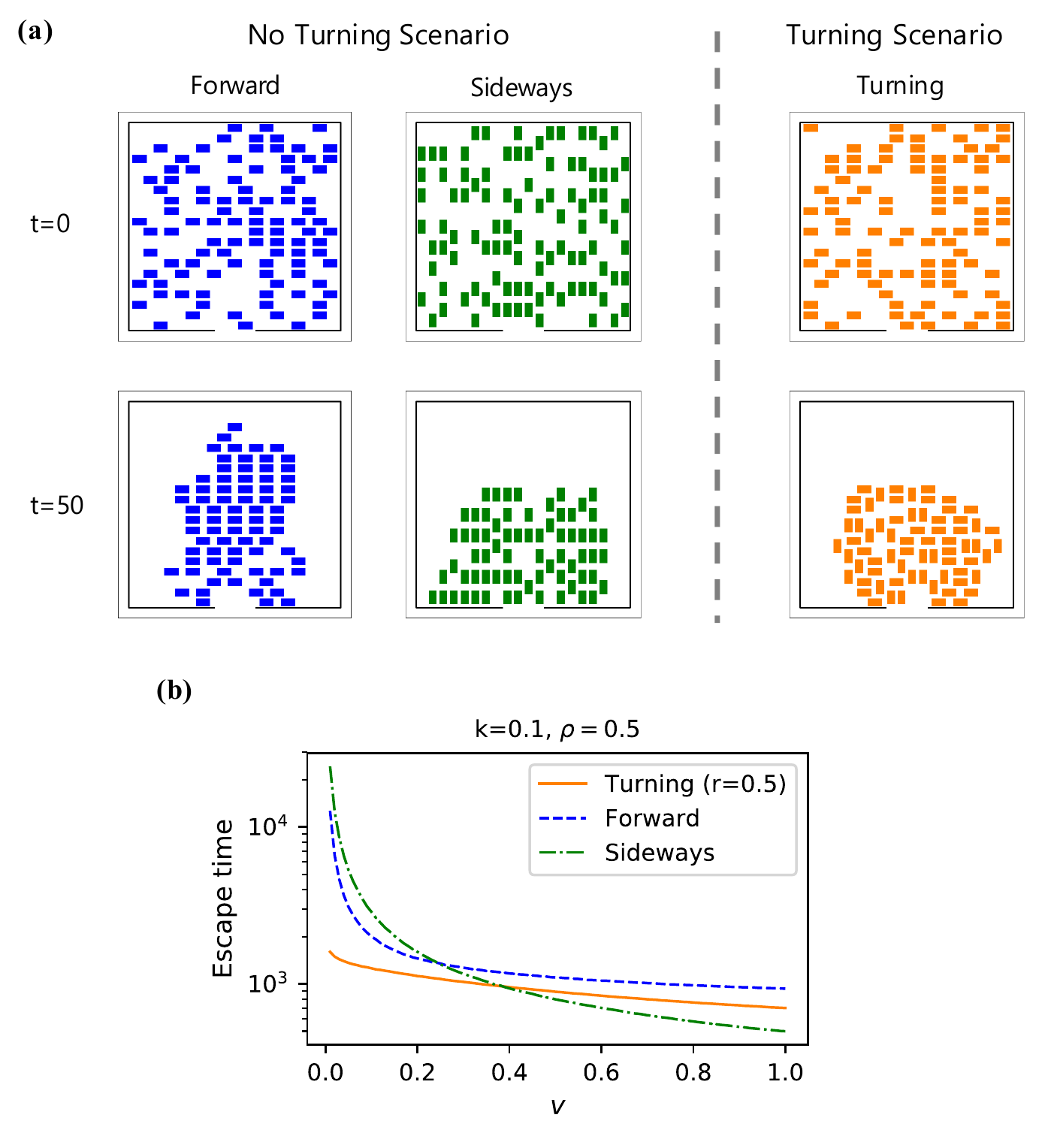}
	\caption{(a) Simulation snapshots ($t=0$ and $t=50$) of three scenarios. A rectangle denotes an evacuee. Forward (shown left) and sideways (shown center) scenarios are treated as the control experiments. In the two scenarios, evacuees do not turn. The difference is the initial body orientation. Turn (shown right) is the scenario we focus on. In the initial setting, evacuees face the exit the same as the forward scenario but they move toward the exit by sometimes turning. We used $\rho=0.5$, $r=0.5$, $v=0.33$, and $L=20$ (Here, small $L$ is piked up for the purpose of snapshots). 
	(b) Escape time in the three scenarios as a function of the speed of sideways movement $v$.}
	\label{fig_fst_explanation}
\end{figure}

To reveal the effect of turning on the escape time, we compared the movement with turning to the one without turning.
As the control experiments (without turning), we set up the two extreme cases, forward and sideways, as shown at the left two in Fig.~\ref{fig_fst_explanation}(a).
In both control experiments, evacuees do not turn.
The difference between forward and sideways scenarios is the direction of the bodies at the beginning.
In the forward scenario, evacuees face the exit while in the sideways scenario, evacuees face sideways to the exit.
The rightmost scenario is our focus where evacuees move toward the exit. Turning is allowed when they move.
In this case, the same as the forward scenario, evacuees face the exit at the beginning.
In all three scenarios, the position of each evacuee is randomly assigned at the beginning (Fig.~\ref{fig_fst_explanation}(a), $t=0$).

We focus on how turning affects the escape time.
Thus, we compared the escape time in the three scenarios.
Figure \ref{fig_fst_explanation}(b) shows the escape time as a function of $v$ (relative velocity of the sideways movement) in each scenario.
The escape time is the total elapsed time from the initial setting to the time that the last evacuee reaches the exit.
We averaged over 100 simulation runs for each data point and the interval between every two data points is 0.01.
Note that the vertical axis is a logarithmic scale.

\begin{figure}[tb]
	\centering
	\includegraphics[width=157.5mm]{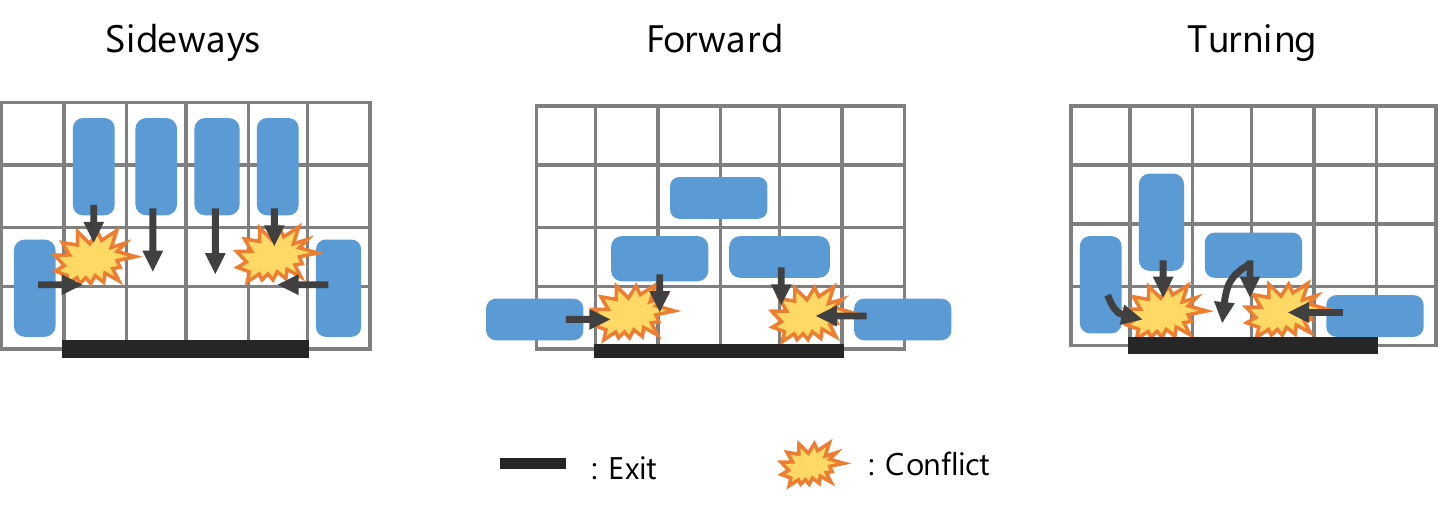}
	\caption{Situations likely to occur around the exit in the three scenarios.}
	\label{fig:fst_conflict}
\end{figure}

As can be observed in Fig.~\ref{fig_fst_explanation}(b), the escape time decreases as $v$ becomes larger.
This is because the increase of $v$ speeds up the sideways movement in any scenario.
If $v$ is equal to or greater than 0.37, the sideways scenario results in the fastest escape time.
We consider that this result depends on the direction that evacuees face near the exit.
Figure \ref{fig:fst_conflict} illustrates the situation likely to occur around the exit.
In the sideways scenario, the escape size of evacuees is smallest to the exit size $D_w$ because evacuees face sideways to the exit.
The size is one against $D_w=4$. Thus, four evacuees can simultaneously escape from the room at the maximum.
Moreover, as shown leftmost in Fig.~\ref{fig:fst_conflict}, even if congestion occurs at the edges of the exit, two center evacuees can still escape from the room.
On the other hand, in the forward scenario, only two evacuees can simultaneously escape from the room because evacuees face the exit.
If the congestion as shown center in Fig.~\ref{fig:fst_conflict} happens, no evacuee can escape from the room.
In the turning scenario, both sideways and forward cases happen. Thus, the result is in between sideways and forward results.
We conclude that, when the relative speed of sideways movement $v$ is large, the sideways scenario results in the fastest escape time because the relative size to the exit is smallest.

Next, we move to the case when $v$ is small ($v < 0.37$) where the turning scenario resulted in the fastest escape from the room.
In this range, although the speed of sideways movement is quite slow, the turning scenario can sometimes avoid this due to turning.
In addition, in that scenario, evacuees can sometimes use sideways movement and more evacuees can escape from the room.
There is another important merit in the turning scenario.
In this case, some evacuees use forward movement and some other evacuees use sideways movement at certain moments.
This makes time differences on the movements on evacuees to the exit.
Thus, evacuees can escape more smoothly from the room in the turning scenario than the other two scenarios.

It is natural to assume that sideways movement is slow.
Actually, the relative speed of sideways movement is 1/3 of the speed of forward movement \cite{Jin2017PhysicaA}.
When $v=1/3$, our result suggests that the turning scenario is best.
Thus, we emphasize that turning behavior can lead to the shortest escape time in a real situation.

\subsection{Optimal value of the turning rate\label{sec:result_opt}}
\begin{figure}[tb]
	\centering
	\includegraphics[width=179mm]{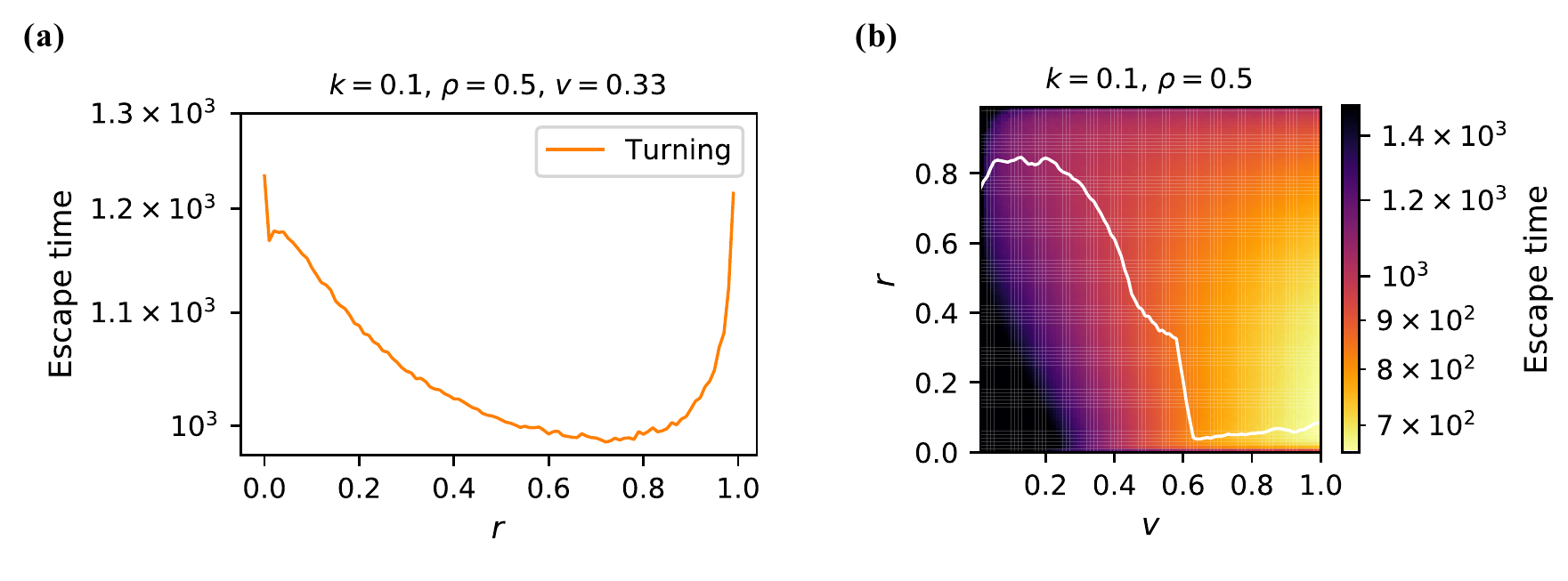}
	\caption{(a) Escape time as a function of the turning probability $r$ in the turning scenario.
	(b) Heatmap of escape time as functions of $r$ and the relative speed of sideways movement $v$. The white line denotes the shortest escape time (optimal value) when $r$ and $v$ values are given. To make the line smooth, we used a central moving average for each point with the window size set to 5.}
	\label{fig:r_cmap}
\end{figure}

So far, we fixed the turning probability $r=0.5$ in the turning scenario.
Here we find the optimal value of $r$ that makes evacuees escape fastest.
Thus, as shown in Fig.~\ref{fig:r_cmap}(a), we changed $r$ while keeping $v=0.33$.
The optimal $r$ must depend on $v$.
We also changed $v$ as shown in Fig.~\ref{fig:r_cmap}(b).

In Fig.~\ref{fig:r_cmap}, we changed $r$ in the range of $0 \leq r \leq 0.99$ with the interval of 0.01.
Each data point is averaged over 100 simulation runs.
Figure \ref{fig:r_cmap}(a) shows a non-monotonic behavior of $r$.
We found that the optimal value of $r$ was 0.72.
Until the optimal value from $r=0.00$, the escape time gradually decreases.
Because we use the relatively slower speed of sideways ($v=0.33$), turning is effective to change the state from sideways to forward.
However, once $r$ is larger than the optimal value, turning is detrimental.
The reason is that when evacuees move sideways, they do not move forward by turning (they can move forward by turning when they are in the forward state).
On the other hand, evacuees can always move forward when they face the exit and do not turn.
In short, by using turning, the average speed of moving forward becomes slow.
Thus, too much turning is detrimental.

Figure \ref{fig:r_cmap}(b) shows how $r$ depends on the speed of $v$.
As the escape time becomes shorter, the color becomes lighter.
The white line implies the optimal $r$ value for each $v$.
This result supports the result in Fig.~\ref{fig:r_cmap}(a).
When $v$ is relatively small, the speed of sideways movement is slow.
Thus, turning from sideways to facing forward is important to escape faster.
However, if $v$ is large, this merit is greatly reduced and the detrimental point of turning becomes large.
Therefore, in that situation ($v > 0.64)$, the optimal $r$ becomes very small ($r \sim 0.1$).

\begin{figure}[tb]
	\centering
	\includegraphics[width=3.28in]{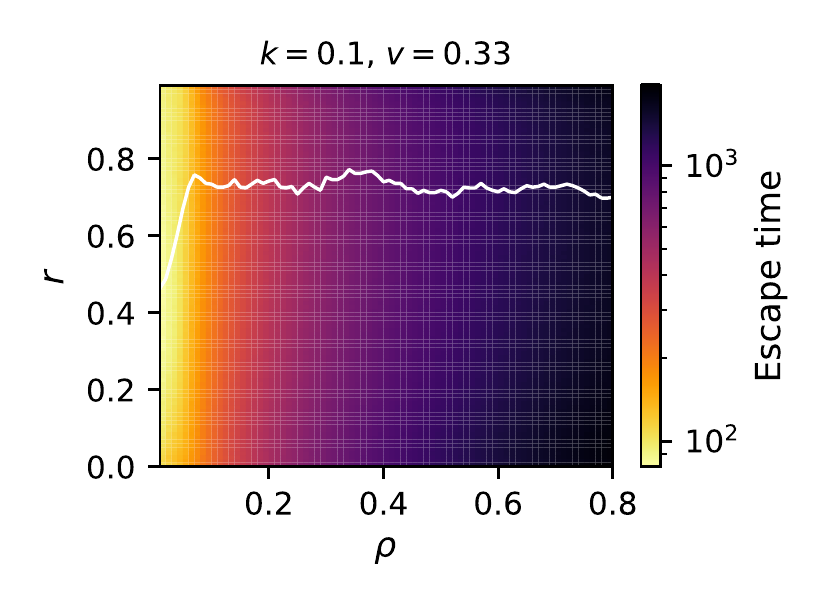}
	\caption{Heatmap of escape time as functions of $r$ and the density $\rho$. The white line denotes the shortest escape time (optimal value) when $r$ and $\rho$ values are given. We used a central moving average for each white point with the window size set to 5.}
	\label{fig:rho_cmap}
\end{figure}

Next, we focus on how the density $\rho$ changes the optimal value of $r$ (Fig.~\ref{fig:rho_cmap}).
We changed $\rho$ in the range of $0 \leq \rho \leq 0.80$ with the interval of 0.01.
The escape time monotonically increases with the increase of $\rho$.
This is obvious because more evacuees need to escape from the room when $\rho$ is increased.
When $\rho$ is very low $\rho < 0.07$, the optimal value $r$ increases around 0.5 to 0.7.
In this situation, congestion rarely happens because the density is very low.
Thus, basically moving forward is the best choice for evacuees to escape fast.
However, if evacuees are not located at a straight position from the exit, they have to move there at some point.
If this happens, evacuees can move to the straight position faster by turning because sideways movement is very slow ($v=0.33$).
This is the reason that some $r$ is needed.
When $\rho > 0.07$, the optimal value of $r$ remains high.
Compared to $\rho \leq 0.07$, a slightly higher $r$ value ($r \sim 0.7$) is needed to avoid a long congestion line to the exit.
However, this effect does not change even if $\rho$ becomes increasingly higher.
Thus, compared to $v$ (Fig.~\ref{fig:r_cmap}(b)), the change of $\rho$ does not affect the effectiveness of $r$ so much.
Finally, the relative speed of sideways movement $v$ has a big impact on the optimal $r$ compared to the density $\rho$.


\section{Conclusions}
In this paper, we studied the effect of turning and sideways movement due to the rectangular size of evacuees on a crowd evacuation.
For that purpose, we developed a cellular automaton evacuation model where evacuees move toward one exit based on the Static Floor Field.
The simulation results showed that the turning behaviors make the escape time from a room shorter.
Moreover, we found the optimal value of the turning rate depending on the relative speed of sideways movement.
These results imply merits and demerits of each movement in a panic situation. 

In the present model, we focused on the three types of typical movements in emergency situations.
We also assumed that evacuees roughly know the location of the exit.
However, in a real panic, some evacuees may not know where the exit is and may act irrationally or unexpectedly which disturbs the evacuation of other people.
Such behaviors may greatly affect the escape time.
One extension of our model is to include such irrational behaviors.
Moreover, in a real panic, dead-lock (almost no one can escapes from the exit) can sometimes happen in a congestion area \cite{XuSong2009BuildEnv}.
It is also important to consider such a case by changing the door size in our model in the future.

Related to this, in our model, we did not consider the difference of personality among evacuees, which also may affect the escape time.
This extension can be done by incorporating a game theoretic perspective to the model as some other evacuation studies have already done.
This is another possible direction in the future.
By understanding human decision-making in emergency situations more deeply, simulation models contribute to develop effective evacuation policies so that people can escape from those situations safely and smoothly.


\appendix

\section*{Acknowledgement}
We thank Takashi Nagatani for his valuable comments to improve this manuscript.


\end{document}